\documentclass{article}

\usepackage{amsmath}
\usepackage{authblk}
\def\ie{$\it i.e.$ }
\def\eg{$\it e.g.$ }

\def\beq{\begin{equation}}
\def\eeq{\end{equation}}
\def\beqn{\begin{equation*}}
\def\eeqn{\end{equation*}}

\begin{document}

\title{Do Black Holes have
Singularities?}

\author[]{R. P. Kerr}

\affil[]{\begin{center}
    University of Canterbury, Christchurch\\
Lifschitz Prof. ICRANet, Pescara 
\end{center}}

\maketitle

\begin{abstract}

There is no proof that black holes contain singularities when they are generated by real physical bodies. Roger Penrose\cite{rp1} claimed  sixty years ago that  trapped surfaces inevitably lead to  light rays of finite affine length (FALL's). Penrose and Stephen Hawking\cite{SH}  then asserted that these must end in actual singularities. When they could not prove this they decreed it to be self evident.  It is shown that there  are counterexamples through every point in the Kerr metric. These are asymptotic to at least one  event horizon and do not end in singularities.

\end{abstract}

\section{History of singularity theorems.}

{\it Note: The word "singularity"  will be used to mean a region or place where the metric or curvature tensor is either unbounded or not suitably differentiable.  The existence of a FALL by itself is not an example of this.}

From 1916 until 1963  the Schwarzschild metric\cite{Schwarz} was  the only known solution   of the Einstein gravitational equations for  the field outside a physically realistic source. At first it was believed that there was a singularity or firestorm around its event horizon but Eddington\cite{Edd} and Finkelstein\cite{Fin}    showed that this was false\footnote{Penrose  used these Eddington-Finklestein coordinates in his 1965 paper\cite{rp1}.}. People's  attention then shifted to the possible curvature  singularity at the centre of Schwarzschild. Was this inevitable  for all real bodies that collapse inside an event horizon? Could it be replaced by a well-behaved, nonsingular matter distribution? This was shown to be impossible if the matter satisfied  various simple equations of state but if the pressure to density relation rises quicker than these then no singularity is required. Also,  how quantum matter interacts with geometry is unknown for such densities.

The Kerr metric\cite{K63} was constructed in 1963,  soon   after the discovery of Quasars. It has a singular source with angular momentum as well as mass, surrounded by two elliptical event horizons. The region between these will be called the "event shell", for the want of a better name.     Objects that enter this are compelled to fall through to the interior. Kerr itself is source-free, "generated"  by a ring singularity at its centre. It cannot be nonsingular since   GR would then admit smooth, particle-like solutions of the Einstein equations that are purely gravitational and sourceless! The ring singularity  is just a replacement for a  rotating star.

The consensus view for sixty years has been that  all  black holes have singularities. There is no  direct proof of this, only the papers by  Penrose\cite{rp1}   outlining a proof that all Einstein spaces  containing a "trapped surface" automatically contain FALL's. This is almost certainly true, even if the proof is marginal. It was then decreed, without proof, that these must end in  actual points where the metric is singular in some unspecified way.  Nobody has constructed any reason, let alone proof for this. The singularity believers  need to  show why it   is true,  not just  quote the  Penrose assumption.

The original Kerr-Schild  \cite{ks1}  coordinates  were deliberately chosen to be  a  generalisation  of Eddington's,  avoiding any coordinate singularities on either  horizon.  It will be shown in Section~\ref{S:Kerr} that  through every point of these spaces there are light rays that are asymptotically tangential to one or other  horizon,    do not have endpoints and  yet their affine lengths are finite\footnote{Counter-examples are the best way to disprove a false conjecture!}. Their tangents are all  "principal null vectors" (PNV's), characteristics   of the conformal tensor. Half of these rays are confined to the event shell. going nowhere near the centre  where the singularities are supposed to be.  Many  have tried to counter these examples  by appealing to the  Boyer-Lindquist extension. This is constructed from a collection of copies of the  separate parts of the original metric, does not include any interior collapsed stars and therefore is known to be singular. It has to be assumed that each interior section contains a star and so one has the same problem as  for Kerr itself.   Furthermore, this extension cannot be formed when a real star collapses: it has nothing to do with physics.  It has not been proved that  a singularity, not just a FALL, is inevitable when an event horizon forms around a  collapsing star. 
We will discuss later why  nonsingular collapsed neutron stars can generate Kerr.

After the First Texas Symposium on Relativistic Astrophysics (Nov.1963), Ray Sachs and I tried to construct an interior solution for Kerr by replacing its ring singularity at $r=0, z=0$  with a finite, non-singular,  interior metric with outer boundary at $r=r_0 > 0$ (say), lying inside the inner horizon. We started by  constructing the Eddington-Kruskal type coordinates that were   independently calculated by Robert Boyer later that year. We used   a  preprint of an outstanding paper by Papapetrou\cite{Pap66} on stationary, axisymmetric Einstein spaces which showed that if these are asymptotically flat with no singularities at infinity then they can be "almost-diagonalised", $\it{i.e.}$ put in a form containing only one off-diagonal term, the  coefficient of $d\phi dt$.   Kerr satisfied all these required conditions. Eliminating the other unwanted components involves  first solving two trivial linear algebraic equations  for the differentials of the new coordinates. Could these be integrated? The crux of Papapetrou's proof is that they can because of two  first integrals which are automatically zero if  the metric is asymptotically flat. This was exactly what was needed to construct these  coordinates in a quite trivial fashion.   The final metric is singular on the two event horizons but it does seem simpler away from those so we hoped that it would help us. However, after ten minutes looking at the resultant metric we realised that calculating such an interior was far more difficult than we expected and  needed us to make assumptions about the properties of the matter inside.  We gave up, cleaned the blackboard, and went for coffee.  We were still convinced that there are many solutions to this problem, some of which may have different inner horizons to Kerr.

The  problem is that there is an infinity of possible solutions but their Einstein tensors do not necessarily satisfy appropriate physical conditions. There have been many such interior solutions calculated since 1963, using various assumptions, but they have  all been ignored because of the false singularity theorems "showing" they cannot exist. Some of these interiors may even be correct! Penrose outlines a proof that if the  star satisfies certain very weak energy conditions and has a trapped surface then it must have at least one FALL. This is true but is little more than the "hairy ball" theorem.

The simplest example of a  FALL  was  calculated  a few days before the  "First Texas Symposium on Relativistic Astrophysics" in November 1963. It lies on the rotation axis between the two event horizons and is asymptotic to each of these. It is what one gets when a torch is shone "backwards" while falling into a black hole down the axis. It does not cross either horizon. This was used at that time to show that the metric has two event horizons, although  I was unable to calculate the general form of these, not knowing of Papapetrou's work until early in 1964. All the examples of  FALL's given in this paper are similarly asymptotic to an event horizon.  They arise because of the interaction between the  light-like Killing vectors that are the normals to the event horizons (and therefore lie inside them)  and the light rays that approach these tangentially, giving converging pencils. These are exactly what  Raychaudhuri\cite{Ray}  studied originally.  His analysis purports to show that a pencil of light rays  satisfying some geometrical and physical conditions will converge at a conjugate point a finite parameter distance away, giving a "singularity". Penrose, Hawking\cite{SH}, Ellis and others have used this to prove their theorems. This will be countered herein by simple examples showing that this point may be at infinity and therefore not attained.

The Kerr metric  contain an infinity of FALL's (two through each point) none of which have terminal points. These  are all  "principal null vectors" (PNV's) of the conformal tensor and are tangential to one or other event horizon at infinity.  None end in singularities, except for Schwarzschild at its centre and Kerr on its singular ring (where $r=a<m, z=0$). These solutions are just replacements for a nonsingular interior star with a finite boundary at or inside the inner horizon. There is a  theorem by Hawking claiming that there are similar light rays in both the future  and at the "Big Bang". We know from observation that matter clumps horrendously forming supermassive black holes, but that does not prove that singularities exist. At best these theorems suggest that black holes are inevitable, which  is almost certainly true: ones as large as 100 billion solar masses have been observed by the James Webb Telescope  in the early universe (Oct.2023).

 As Einstein once said, "{\it General Relativity is about forces, not geometry}". This may be a simplification but it is a very useful one.  The Kerr solution can  be used to approximate the field outside a stationary, rotating body with mass $m$, angular momentum $ma$, and radius  larger than $2m$. The best example is  a fast-rotating neutron star too light to be a black hole. How accurate is this metric? Probably better than most! If $R$ is an approximately radial coordinate  then the rotational and Newtonian "forces" outside the source drop off like $R^{-3}$ and $R^{-2}$, respectively\footnote{Calculations by the author used  the corrected EIH equations in the late fifties to show this is accurate for slow moving   bodies at large distances (and reasonable elsewhere)}. Clearly, spin is important close in but  mass dominates  further out.  These are joined by   "pressure"   near the centre where the others vanish. Most, probably all, believe this "standard model" is nonsingular for neutron stars\footnote{Outside the Earth centrifugal force plays a minor role but is still important for  sending  rockets into space. That is why the launch sites are chosen as close to the equator as possible. After the initial vertical trajectory they  travel east with the Earth's rotation rather than west against it. 
}, but not for black holes. Why the difference? The actual density can even be lower for a very large and fast rotating black hole interior.

Suppose a  neutron star is  accreting matter, perhaps from an initial supernova.  The  centrifugal force can be comparable to   the Newtonian force near the surface\footnote{If the body rotated too quickly then the surface would disintegrate. This puts a lower limit on the possible size of the star.}, but further out there will be a  region    where it drops away and   mass dominates. It can  be comparatively easy to launch a rocket from the surface, thanks to the slingshot effect; further out it will require a  high velocity and/or acceleration to escape from the star.   This intermediate region  will gradually become a no-go zone as the mass increases and the radius decreases, \ie  an event shell and therefore   black hole  forms. Why do so many believe that  the star inside must become singular at this moment? Faith, not science! {\it Sixty years without a proof, but they believe!}. Brandon Carter calculated the geodesic equations inside Kerr. showing that  it is possible to travel in any direction between the central body and the inner horizon.. There is no trapped surface in this region, just  in the event shell between the horizons.

The work of David Robinson and others shows that a real black hole will have the Kerr solution as a good approximation to its exterior but a physically realistic, non-vacuum, non-singular interior. Since these objects are also accreting, both horizons of Kerr should be replaced by apparent horizons. As the black hole stops growing,   Kerr is likely to be a closer and closer approximation outside the inner horizon. The singularity theorems do not demonstrate how (or if) FALL's arise in such environments but that of Hawking claims that these must always form in our universe, given that { \it almost-closed} time-like loops do not.\footnote{Hawking originally claimed, when visiting UT for a weekend, that closed loops were the alternative. I said in a private conversation to Hawking and George Ellis that after thinking about it over the weekend I could not quite prove this, just that "almost-closed" loops were the alternative. Steven subsequently changed his paper to agree with this. A different  name is given in Hawking and Ellis\cite{HE} and attributed to me.}  It is probably true that the existence of FALL's show that horizons exist and that these contain black holes. Proving this would be a good result for a doctoral student. There are indications that these  are inevitable.  Astronomers are now seeing them more and more.  Matter clumps!

Several people have said  "What about the analytic extensions of Kruskal and Boyer-Lindquist?", implying that the singularities could be there.   These  extensions may be analytic, but at best they are constructed using copies of the original spaces together with some fixed points. These will be nonsingular inside each copy of the original interior if the same is true inside the original Kerr and therefore the extensions are irrelevant to the singularity theorems. Anyone who does not believe this needs to supply a proof.  They  are all physically irrelevant since real black holes start at a finite time in the past with the collapse of a star or similar over-dense concentration of matter, not as the white hole of the  Kruskal or Boyer-Lindquist extensions. They continue to grow for ever, perhaps settling down to some final size (or evaporate if the latest proof of Hawking's theorem is  true!).\\ 
{\it "Science is what we have learned about how to keep fooling ourselves." Richard Feynman.}

\section{Affine parameters}\label{S:Affine}

This short section is the crux of the argument that the singularity theorems are proving something different to "singularities exist!". The reason that so many relativists   have assumed  that Raychaudhuri's theorem proves that bounded affine parameter lengths lead to singularities is that they have confused affine   with geodesic distance.  These are very different concepts. Geodesic parameters are defined by a {\it first-order} differential equation,
\begin{equation*}                                      \frac{ds}{dt}  = \sqrt{g_{\mu\nu}\frac{d x^\mu}{dt}\frac{dx^\nu}{dt}},\qquad \longrightarrow \qquad s=s_0 + C,
\end{equation*}
where $t$ is an arbitrary parameter along the ray,  perhaps a time coordinate, $s_0$ is a particular solution, and $C$ is an arbitrary constant. 

This does not work for  light rays where $ds = 0$. Its replacement, affine "distance", $a(t)$, is defined by a {\it second order} differential equation instead. Since the acceleration is proportional to the velocity for a geodesic,  \begin{equation}\label{E:affine:t}
     \frac{d^2 x^\mu}{dt^2} + \Gamma^{\mu}_{\alpha \beta}\frac{dx^\alpha}{dt}\frac{dx^\beta}{dt}  =  \lambda(t)\frac{dx^\mu}{dt}.      
\end{equation}
where  $\lambda$    is a function along the curve. The parameter $t$ can be replaced by a function $a(t)$  chosen to eliminate $\lambda$,
\begin{equation}\label{E:affine:s}
   \frac{d^2 a}{dt^2} = \lambda \frac{d a}{dt} \quad \Longrightarrow \quad
\frac{d^2 x^\mu}{da^2} + \Gamma^{\mu}_{\alpha \beta}\frac{dx^\alpha}{da}\frac{dx^\beta}{da}  =  0.
\end{equation}
The  tangent vector, $\frac{dx^\mu}{da}$, is then parallely propagated along the ray. The general solution for $a$ is 
\begin{equation}\label{E:affine:a}
    a=Aa_0 +C.
\end{equation}
where A and C are arbitrary constants and $a_0$ is a particular solution.   This transformation is affine; $a$ is called an affine parameter.  The crucial difference between the two parameters, $s$ and $a$, is that if $\lambda$    is a constant in eq.(\ref{E:affine:s}) then  $a_0 =e^{\lambda t}$ and {a(t)} is bounded at either 
$+\infty$  or $-\infty$. 
This is also true if $\lambda$ is bounded away from zero, $|a(t)| > B_0 > 0$ 
where $B_0$ is a nonzero constant. This has nothing to do with singularities. 

Suppose that $k^\mu$ is a Killing vector with an associated coordinate $t$,
\begin{equation*}
    k_{\mu ; \nu} +  k_{\nu ; \mu} = 0,\qquad k^\mu \partial_\mu = \partial_t,
\end{equation*}
and that it is  also a light ray along one of these curves. Multiplying by $k^\nu$,
\begin{equation*}
    k^\nu k_{\nu;\mu} = 0 \quad \longrightarrow \quad k^\mu _{;\nu} k^\nu = 0, 
\end{equation*}
and so this particular curve is also geodesic and the $t$-parameter, {\it or any affine function of it}, is affine!

 We will see that the normals to each of the event horizons of Kerr and Schwarzschild are such light rays, PNVs lying  in the horizons. They are  invariants of the symmetry group and  are constant multiples of $\partial_t$. Each of these  is a light-like vector and is itself a  Killing vector. Their affine parameters are  exponential functions of the time parameter, $ Ae^{Bt} + C$. Choosing $C=0$,
\beq\label{E:Sym}
a(t) = Ae^{Bt},
\eeq
where $(A, B)$ are  constants, and so it vanishes at one or other end unless $B=0$. This has nothing to do with singularities.

\section{Schwarzschild and Eddington.}

When Karl Schwarzschild\cite{Schwarz} first presented his solution (referred to as S) for a spherically symmetric Einstein space,
  \begin{equation*}
ds^2 =  -(1-\frac{2m}{r})d{t_S}^2 + (1-\frac{2m}{r})^{-1}dr^2 + r^2 d\sigma^2, \qquad
 d\sigma^2 = d\theta^2 + sin^2\theta d\phi^2,\\      \end{equation*}
it appeared to have two singularities. The first was at its centre where  the curvature tensor was  infinite, the second at the event horizon, ${r = 2m}$. For several years it was thought that the latter was real and that there was a firestorm on this surface. Eddington\cite{Edd} and  Finklestein\cite{Fin}  showed that this was false by writing the metric in different coordinate systems where the only singularity was at the centre. They also showed that any object that crossed the horizon would quickly fall to this "point". 

 The time-coordinates, $t_-$ and $t_+$, respectively, of the two forms of Eddington,
 ingoing $E_-$, a "black" hole, and outgoing $E_+$, a so-called "white" hole, are related to Schwarzschild time, $t_S$, by 
 \begin{subequations}\label{t_trans}
    \beq\label{t_transa}
   t_- =t_S - 2m\ln| r - 2m |,\qquad t_+ = t_S + 2m\ln| r - 2m |, 
     \eeq 
     \beq\label{t_transb}
    t_+ = t_- + 4m \ln| r-2m |,
     \eeq
 \end{subequations}
where we use the   subscripts, $(S,-,+)$ on the time coordinates to distinguish them.  The other three coordinates $(r, \theta, \phi)$   do not require indices because they do not change.

The two Eddington metrics have the Kerr-Schild form\footnote{When an equation contains $\pm$ or $\mp$ signs the top group give one equation, the bottom another.},
\begin{equation*}
    ds^2  = d s^2_{0\pm}  + \frac{2m}{r}(k_{\pm\mu} dx^\mu )^2,
\end{equation*}
 where  the first term is the corresponding Minkowski metric,
 \begin{equation*}
   ds_{0\pm}^2  = dr^2  + r^2 d\sigma^2  -dt_\pm^2,
\end{equation*}
and  the (${k_\pm ^\mu, k_{\pm\mu}}$) are light rays for both the background spaces and the full metrics,

\begin{equation}\label{PNV_m}
k_{\pm} = k_{\pm\mu} dx^\mu = \pm dr -  dt_\pm, \qquad  
\bf{k}_\pm= k_\pm^\mu \partial_\mu =    \pm  \partial_r + \partial_{t_\pm}.
\end{equation}
The transformations in (\ref{t_transa}) are both singular at the event horizon, $ r=2m $, but the two metrics themselves are analytic. That is also true for the  appropriate radial light rays, $\bf{k}_\pm $,  that point inwards for $E_- $ and outwards for $E_+$.  
Since the second PNV, $\bf{k}_\pm^*$  say, in one coordinate system is the first one  in the other, $\bf{k}_\mp$, it is easily calculated using (\ref{t_transb}),
 \begin{equation*} 
 k^*_{\pm\mu} dx^\mu = \pm \frac{r-2m}{r+2m} dr -dt_\pm, \qquad {\bf k^*_\pm} = k^{*\mu}_\mp \partial_\mu = \pm\frac{r-2m}{r+2m}\partial_r +  \partial_{t_\pm}.
 \end{equation*}
For a black hole, both $\bf{k}_-$ and $\bf{k}_-^*$  point inwards inside the event horizon at $r=2m$. Outside this $\bf{k}_-$ points inwards whilst $\bf{k}_-^*$ points outwards. The two Eddington  metrics are identical if one allows a simple inversion of time, $t_+ \longleftrightarrow -t_-$ but  this inverts the orientation. Since physical metrics are always oriented, this is not permissable.

\emph{ NOTE: We can think of $(K_\pm, K_S)$ as three separate spaces or three coordinate systems on the same space. In the second case, at least
two of the coordinate systems are singular.  If we start with a Black Hole then $K_-$ is nonsingular, the other two are singular.}

The two families of light rays are the characteristic double "principal null vectors" (PNV) of the conformal tensor and  are  both geodesic and shearfree.    Neither ray crosses the event horizon in the original Schwarzschild coordinates but   $\bf{k}_-$does in $E_-$ coordinates whilst the other, $\bf{k}^*_- $, is asymptotic to it as $t_- \rightarrow  -\infty $. There are two PNV's at each point of the horizons themselves. One goes  through  but the other lies in  the horizon and is its normal, $\partial_t$, at that point.  None of this is new. It has been known for almost a century.

The second set of PNV's are asymptotic to the event horizon as $t_- \rightarrow -\infty$ for a black hole and as $t_+ \rightarrow +\infty$ for a white hole.  In both cases  the affine parameter $r$  is necessarily bounded as the PNV approaches the appropriate horizon\footnote{We will see later that the second PNV in Kerr, $k_\pm^*$, is  asymptotic on both sides to the inner horizon at $t=+\infty $ and  to the outer horizon at $t=-\infty. $ It is  a FALL ray between the horizons.}. Since the metrics are stationary this is an example of the predictions of section \ref{S:Affine}. This contradicts the basic assumption that ALL  singularity theorems are based on. The only reason that it is assumed that these rays must end at a singularity is so that these "theorems" can be proved. This includes Hawking's, Penrose's and all other similar theorems for black holes and the "big bang". They are built on a foundation of sand. We will leave this for the moment until we have introduced the Kerr metric where the examples are even clearer.

{\it"The human brain is a complex organ with the wonderful power of enabling man to find reasons for continuing to believe whatever it is that he wants to believe."}-Voltaire.

\section{The Kruskal Extension of Schwarzschild}

Many have said to the author "What about the Kruskal-Szekeres\cite{Krus,Szek}  extension?" as if this makes a difference to any singularities. The original treatment of this starts with the singular Schwarzschild coordinates, $"S"$, and then uses a singular transformation to generate the  Kruskal coordinates, $"K"$. This  has been used in lectures for decades but the resulting metric is itself singular on the horizon where its determinant behaves like $\sqrt{r-2m}$.  Instead of this, we will use the more recent approach to show  that the proper  Kruskal metric is  an analytic extension of Eddington, rather than Schwarzschild.

The two coordinates $(\theta, \phi)$ are retained but the other two $(r,t)$ are replaced by  $(U,V)$ that are constant along the ingoing and outgoing PNV's, respectively. For simplicity, we will assume we start with a black hole with ingoing coordinates, $E_-$. but will omit the $\pm$ sign on the metrical components. Also, we will use units where $2m=1$ so that the horizon is located at $t=1$, not $t=2m$. This makes the calculations more readable.  Starting with Eddington,
\begin{align*}
    ds^2 \, &= \, dr^2-dt^2 +\frac{1}{r}(dr + dt)^2 \, = \, (dr+dt)(dr - dt +  \frac{1}{r}(dr + dt))\notag\\ 
    &= \,\frac{r-1}{r}d(r+t)(\frac{r+1}{r-1}dr - dt)\, =\, \frac{r-1}{r} du dv,           
\end{align*}
where $(u,v)$ are given by
\begin{equation*}
    u =r+t, \qquad v=r+2 sgn(r-1) ln|r-1| - t 
\end{equation*}
Each of the coordinates, $(u,v)$, is  constant along the appropriate PNV but  $v$ is useless near the event horizon at  $r=1$ because it is singular there. They need to be replaced with functions, $(f(u),g(v))$, that are positive and at least three times differentiable at the horizon. The standard choice is to exponentiate them,
\beq\label{UandV}
        U =  e^{u/2} =  e^{\frac{r+t}{2}},  \qquad   V = e^{v/2} = e^\frac{r-t}{2}(r-1),
\eeq
an analytic transformation. This leads to the Kruskal-Szekeres  metric,
    \beq\label{ksa}
            ds^2 =  \frac{4}{r} e^{-r}dU dV  +r^2 d\sigma^2 \Longrightarrow  \frac{32m^3}{r} e^{-r/2m}dU dV  +r^2 d\sigma^2,
    \eeq
where we have reintroduced the  $2m$ factors!  Each new coordinate is zero on  one of the "perpendicular"  event horizons of Kruskal. The  metric coefficients in \eqref{ksa} are functions of $r$ alone so we only need to calculate this as a function of $U$ and $V$. From eq.\eqref{UandV}, 
\beq\label{UV}
    UV = (r-1)e^r = g(r)\quad (say), \qquad  \frac{dg(r)}{dr} = re^r.
\eeq
From a standard theorem in analysis this can be solved for $r$ as an analytic function of $UV$ in the connected region where $r$ is positive and the derivative of $g(r)$ is nonzero, \ie away from the real singularity at $r=0$. This is even true in the complex plane, except for the branch point at the origin. This allows the definition of the coordinates to be extended to negative values giving the proper Kruskal metric.  This has two horizons, ($U=0$ or $V=0$, where $r=2m$) which can only be crossed in one direction because  "time" flows is a unique direction. The lines of constant $r$ are the hyperbolae where $UV$ is  constant. Both the Jacobian matrix for the map from $(r,t)$ to $(U,V)$ and its inverse are analytic so the map from Eddington to Kruskal coordinates is analytic. This is not true when going from Schwarzschild to Kruskal.

There is another  major difference between Eddington and Kruskal. The former is stationary, \ie independent of time, but the later has a fixed point at its center and is invariant under a boost symmetry,
\begin{equation*}
    U \rightarrow \lambda U, \qquad  V \rightarrow \lambda^{-1} V.    
\end{equation*}

Kruskal is a valid extension but has no real physical significance. The second singular region is usually thought of as a white hole, generated by a nonsingular time-reversed object replacing the singularity at $r=0$. There is no more likelihood of a singularity in this region than there is in Eddington itself. If one believes in nineteenth century equations of state, regardless of pressure, then anything is possible.\footnote{What this all means, I have no idea! I do not believe anyone else does either since the behaviour of quantum matter at such  extreme pressure is unknown.} Also, black holes form in our universe when matter accumulates into clumps that are too massive and/or dense. They do not start as white holes \`a la Kruskal. 

Suppose that there is a nonsingular spherically symmetric star, $S$, at the center of Eddington, and suppose it is bounded by $r=r_0$. All the incoming PNV's are radial geodesics, passing through the central point and dying on the opposite side of the star. Since the radius is an affine parameter on each ray the total affine length from crossing the event horizon is less than $4m$, a finite number. The possible infinite curvature at the centre has nothing to do with the singularity theorem, just the claimed physics of the star. The affine parameters are irrelevant.

  \section{The Kerr metric}\label{S:Kerr} 

The conformal tensor for an empty Einstein space has four special light rays,  its principal null directions (PNVs). These can be thought of as "eigenvectors" of the conformal tensor.  They generate four congruences. When two coincide everywhere the space is  called "algebraically special". The corresponding congruence is then both geodesic and  shearfree.   Furthermore, the converse is true. If the space has such a special congruence of  geodesic and shearfree light rays then, from  the Goldberg-Sachs theorem\cite{gs}, these are repeated eigenvectors of the curvature tensor. Both the two physically interesting solutions of Einstein's equations known in the first half of the twentieth century, Schwarzschild and plane fronted waves, are algebraically special.   The PNV's  coincide in pairs in the first (type D) and all four coincide 
in the second (type N).

In 1962 Robinson and Trautman\cite{rt} constructed all algebraically special spaces where the double PNV is hypersurface orthogonal, {\it{i.e.}} a gradient. This was the most general solution known at that time but, although it is both very elegant and has many interesting properties, it did not lead to any new star-like solutions. 
Several groups tried to generalise this work to allow for a rotating double PNV. These included Newmann, Unti and Tamborino\cite{nut} who claimed in 1963 that the only new metric of this type was NUT space. Neither Ivor Robinson nor myself believed this result.  The Robinson-Trautman metrics should have been included in their solution as they are a  special case but they were not. Curiously enough, others did believe their results and there was a lot of effort put into the this metric. When I  finally saw a preprint of their paper I flicked through it until I found an equation where the derivation ground to a halt. They had calculated one of the Bianchi identities twice, did not recognise this, made several  mistakes in the numerical coefficients, and got inconsistent results.\footnote{Even fifty years later, Newmann still did not understand where they went wrong. The major problem was that they used the Newmann-Penrose equations where the components of the connexion lack numerical indices. This meant that it was difficult to check that each term in an equation had the correct "dimensions" without the aid of modern computer algebra. Many other people, including myself, were using similar systems but retaining numerical indices on the connexion coefficients.}

 Ignoring the rest of their paper, I set about calculating all algebraically special metrics,  \ie 
 empty Einstein spaces with a double PNV which might not be  hypersurface orthogonal. This led to a set of canonical coordinates, a generalisation of those used by Robinson and Trautman in their seminal work and therefore of those used by Eddington. The first, and most important, of these  was an affine parameter, $r$, along the rays. The dependence of the metric on this was fairly easily calculated (An early and accurate example of the "Peeling Theorem") so that the remaining field variables  were functions of the other three coordinates alone. Unfortunately, the final  equations were not integrable and nobody has been able to  simplify them further without assuming extra conditions.

  The Kerr metric  was discovered in 1963 \cite{K63,wvs} by imposing the following series of conditions on an empty  Einstein space, 
\begin{enumerate}
    \item It contains a field, $k^\mu$, of geodesic and shearfree light rays through each point. The Goldberg-Sachs theorem  proves that this is equivalent to demanding that they are repeated eigenvectors of the conformal tensor. This  gave a set of five partial differential equations which were inconsistent unless an endless chain of further integrability conditions were satisfied. These led nowhere and so  three more  simplifications were imposed in sequence,
    \item It is stationary, i.e. independent of time.  This helped but the equations were still intractable.
    \item It is  axially symmetric. Much better! 
    \item Finally, it is asymptotically flat.
\end{enumerate}
The symmetry conditions reduced the problem to solving some simple, ordinary differential equations\footnote{Andrzej Trautman told the author recently that in the early sixties he set a graduate student the problem of calculating all such Einstein metrics  by starting with the known  field equations for stationary and axisymmetric metrics,${\it i.e.}$ conditions (2,3), and then imposing (1).  This should have led to the Kerr metric but  Andrzej said that mistakes were made and nothing came of it.}. The final assumption, (3), eliminated all possibilities (including NUT space) except for the two parameters, $m$ and $a$, of the Kerr metric. 

Using the original coordinates \cite{K63},
\begin{subequations}\label{ks}
   \beq 
   ds^2 =  ds_{0}^2 + \frac{2mr}{\Sigma} k^2,\quad k =   dr + a\sin^2\theta\ d\phi + dt,\label{E:ks1}
   \eeq
   \beq
    ds_0^2  = dr^2 + \Sigma d\theta^2  
 + (r^2 + a^2)\sin^2\theta d\phi^2 + 2a\sin^2\theta d\phi dt  -dt^2,\label{E:ks2} 
    \eeq\beq
   \Sigma = r^2 + a^2 \cos^2\theta.\label{E:ks3}
   \eeq
\end{subequations}
where the light ray $k$ is a PNV and $ds_0^2 $ is  a version of the Minkowski metric exhibiting the canonical Papapetrou form (Only one off-diagonal term) for metrics satisfying the three conditions above! \\[.6em]
{\it NB: The coordinate $r$ is an affine parameter along a lightray, $k$, when the underlying space is algebraically special and $k$ is a double PNV.}\\[.6em]
This was the first example of the Kerr-Schild metrics\cite{ks1,ks2} which are defined to have the same form as in eq.\eqref{E:ks1}. The sign of $a$ is flipped from that in the original 1963 paper because of my confusion over which direction an axial vector should point in! When $a=0$, the coordinates $(r, \theta, \phi)$  are just spherical polars  in Euclidean space and the metric reduces to Schwarzschild in Eddington coordinates. The  transformation\cite{K63}
\beqn
    x+iy = (r+ia)e^{i\phi}{\rm sin}\theta ,\quad  z = r{\rm cos}\theta,
\eeqn
gives the Kerr-Schild form in more obvious coordinates,
\beq\label{E:KSform}
\begin{split}
    ds^2 = d&x^2 + dy^2 + dz^2 - dt^2 + \frac{2mr^3}{r^4+ a^2z^2}[dt +\frac{z}{r}dz\\
    &+ \frac{r}{r^2+a^2}(xdx+ydy) + \frac{a}{r^2+a^2}(xdy-ydx)]^2.
\end{split}
\eeq
The surfaces of constant $r$ are   confocal ellipsoids of revolution,
$$
    \frac{x^2+y^2}{r^2+a^2} + \frac{z^2}{r^2} = 1.
$$
A simple calculation  shows that the vector $k^\mu$ is a  geodesic in the underlying Minkowski metric as well as the full metric. This is true for all Kerr-Schild metrics. Those PNV's that lie in the equatorial plane are tangential to the  central ring,
\beq
    r=0\quad \longrightarrow \quad x^2+y^2 = a^2,\quad z=0.
\eeq
The rest all pass through this  to a second nonphysical sheet.  The metric is nonsingular everywhere except on this ring. 

This  singularity   generates   the Kerr metric. It must be replaced by an actual rotating body such as a neutron star to construct a physical solution where the central ring and second sheet disappear and the metric is nonsingular. What about the Penrose theorem? We will see that there are plenty of FALL's tangential to the event horizons inside both the event shell and the inner horizon. Also, there is no trapped surface inside the latter to affect the metric of the star.  There is no singularity problem when the ring is replaced by an appropriate star! 

We will  discuss the complete set of PNV's in the appendix but it is simpler to restrict the discussion here to  those  on the rotation/symmetry axis. All others behave exactly like the axial ones. They are asymptotic to the outer event horizon as $t\rightarrow-\infty$ and to the inner horizon as  $t\rightarrow +\infty$, each from both sides. This is amplified in the appendix.
The axial ones are constructed by  calculating the metric  and finding its roots. This will give both the incoming and outgoing light rays,
\beqn
    ds^2 = - dt^2 + dr^2 + \frac{2mr}{r^2+a^2}(dr+dt)^2 = 0,
\eeqn
and so  $\frac{dr}{dt} = -1$ for the incoming geodesic.   For the other,
\beqn
    \frac{dr}{dt} = \frac{r^2 - 2mr + a^2}{r^2 + 2mr + a^2}.
\eeqn
This geodesic cannot cross either horizon as its  radial velocity is zero there. 
Since the RHS is negative between these, both PNV's are pointing inwards in this region. The "fast" null geodesics continue straight through both horizons but the "slow" one is asymptotic to the outer horizon as $t \rightarrow -\infty$ and to the inner horizon  as $t\rightarrow +\infty$. It penetrates neither. It is compelled to move inwards between the horizons and outwards otherwise. Since r is an affine parameter for both these light rays, the affine length of the slow geodesic between the horizons is $2\sqrt{m^2 - a^2}$, a finite quantity.  This is a simple demonstration of what was discussed in section \ref{S:Affine}, contradicting the assumption that null geodesics of finite affine length must end in singularities. The same thing happens to this slow ray as it approaches either horizon from outside the "event shell". It cannot cross them.

What about the two fast incoming geodesics on the axis (one from each end)? These are the rays that Penrose is working with.  Since Kerr has no  interior body they are compelled to pass through the central ring singularity into the other nonphysical  branch of Kerr. If the metric is generated by an axially symmetric and nonsingular neutron star or similar ultra-dense body (whose surface is probably ellipsoidal, $r=r_0$.) then the two incoming axial light rays  will pass through it and swap places on the other side. This means that the fast geodesic coming in will become  the slow one going out  and   be asymptotic to the inner horizon on the opposite side\footnote{This is  an example of exactly what Penrose attempts to prove. If the body is nonsingular then there are FALL's.}. Its affine parameter, $r$, is bounded. Light rays   can approach this horizon (Cauchy surface?) from inside but cannot cross it.

In a truly remarkable paper,  Achilles Papapetrou\cite{Pap66}   discussed stationary and axisymmetric Einstein spaces where the sources are localised and the metric is asymptotically flat. Assuming these conditions, he proved that there is a coordinate system with only one off-diagonal component in the metric, the coefficient of $d\phi dt$. Furthermore, this can be found by solving two simple linear {\it algebraic} equations! In early 1964 Ray Sachs and the author decided to calculate an interior solution for this metric. We believed that the singularity in the centre is not real and that there must be many nonsingular interior "neutron star" metrics that could replace it. Since we had a preprint of Papapetrou's paper we put the Kerr metric into his canonical form.
The covariant form of the metric, $ds^2$, is then  a sum of squares of a suitably weighted orthonormal basis,
\begin{multline}   
       \qquad ds^2 = \frac{\Sigma} {\Delta}dr^2   -\frac{\Delta}{\Sigma}\left(dt_s + a \sin^2\theta \,d\phi_s \right)^2,\qquad\\+ \Sigma d\theta^2  +  \frac{sin^2\theta}{\Sigma}\bigl( (r^2+a^2)d\phi_s - adt_s \bigr)^2
\end{multline}
We stared at this metric for a very short time,  gave up and went for coffee.   The problem is that there are too many possible interior solutions, the same as for regular neutron stars. Physics is needed, not just mathematics! Note that at no point did either of us consider that the interior body was singular.

Unlike the  Kerr-Schild coordinates, the Boyer-Lindquist ones are singular when $\Delta = 0$.   One important advantage of them  is that they make it easy to identify  the event horizons and the two PNV's. The former  are the two ellipsoids  where  $\Delta = 0$. The metric around these surfaces is nonsingular in Kerr coordinates, as in Eddington, but is singular in BL, as in Schwarzschild.

It  is shown in the Appendix that there are two families of characteristic light-rays in Kerr. These are  tangential to a PNV at every point. There  a "fast" one going in unimpeded and a "slow" one trying to get out\footnote{This is probably true for all rotationally symmetric black holes, whether  stationary or not.} but stalling on the horizons. In the original coordinates their contravariant forms are

\begin{align}
    {\bf k}_- &= \partial_t -  \partial_r \quad\Longrightarrow \quad\frac{d r}{dt} = -1,\\
   {\bf k}_+  &= \Delta{\bf k}_-   + (4mr + 2a^2 sin^2 \theta) {\bf \partial}_r,
\end{align}
which shows that  $\bf{k}_+$ lies on each event horizon when $\Delta(r) = 0$ and is  parallel to  its  normal, ${\bf \partial}_r$.
 From  section~\ref{S:Affine},   whenever ${\bf \partial}_t$ is a Killing vector on a light ray then any affine parameter on the ray is an exponential function of $t$. For Kerr, it approaches a constant as $t\longrightarrow +\infty$ on the inner horizon  or as $t\longrightarrow -\infty$ on the outer horizon. These affine parameters can be chosen to be constant on each horizon so that $a$ is a smooth function throughout.

We assume that there is an axially symmetric, smooth,  nonsingular star-like body inside the inner horizon with surface $r=r_S$. Consider a "fast" axial lightray falling into the black hole from outside.  It  moves down the axis, through both horizons, through the star and finally finishes on the other side as a "slow" ray. Does it match up with a PNV coming in from the other side? This is unlikely since there is only two such characteristic light rays at each point in the empty space outside the star, none within. Does at least one (The axial one?) line up? If the Penrose theorem is true, then yes. If not then that theorem  can possibly be modified to show that one incoming light ray is asymptotic to the inner horizon and therefore a FALL.

\section{Conclusions}

The fact that there is at least one FALL in Kerr, the axial one, which does not end in a singularity shows that there is no extant proof that singularities are inevitable.   
The boundedness of some affine parameters has nothing to do with singularities. The reason that nearly all relativists  believe that light rays whose  affine lengths are finite must end in singularities is nothing but dogma\footnote{My experience from listening to graduate students discussing research papers is that they almost always give the mathematics a very cursory glance. This is also true for many professionals. Life is short and they are understandably more interested in physics.}. This is the basis for all the singularity theorems of Hawking, Penrose and others and so these are at best unproven, at worst false. Even if they were true then all they would   prove is that at least one light ray from the outside is asymptotic to an event horizon and is a FALL but one might have to wait for an infinite time to confirm it  for  accreting black holes. 
Proving this  would make a good initial problem for a mathematically inclined doctoral student.

The author's opinion is that gravitational clumping leads inevitably to black holes in our universe, confirming what is observed,  but this does not lead to singularities. It is true that there are "proofs" that the curvature of a non-rotating one is infinite at its central point.These all assume that matter is classical and that it satisfies whatever nineteenth century equation of state the proponents require to prove whatever it is that they wish to prove. Equations of state assume that all variables, such as pressure and volume, occur in the simplest algebraic fashion. This may be true for the low density laboratory or engineering experiments but perhaps  not at black hole densities. The author has no doubt, and never did, that when Relativity and Quantum Mechanics are  melded it will be shown that there are no singularities anywhere. When theory predicts singularities, the theory is wrong!

There are no event horizons when $a>m$ for a Kerr metric   since there are no real roots of $r^2-2mr+a^2 = 0$. It still has a singular ring, radius a. The metric would need either a mass m rotating at the velocity of light at this radius (impossible!) or an actual star with greater radius  and lesser velocity at its equator. Hardly anybody  believes that real stars  contain singularities (Penrose states this as a principal,  counterbalancing his edict that all black holes have singularities!) and so it must be  that centrifugal force combined with internal pressures can overcome the "Newtonian attraction" inside such very fast rotating stars. Also,
the inner region of Kerr allows movement  both inwards and outwards towards the inner horizon, just like the neighbourhood of a regular star. As a star shrinks its centrifugal forces rise rapidly. There is no known reason why there cannot be a fast rotating nonsingular star inside the horizons generating the Kerr metric outside. There is no published paper that even claims {\it to prove} that this is impossible and yet so many believe  "All black holes contain a singularity.".

 The secondary PNV's in Kerr, the $\bf k_+$, do not start at a Cauchy surface outside since they  are tangential to the outer horizon in the past. They are not  counterexamples to what Roger claims. We need to look at the other set of light rays, the $\bf k_-$.  These can start on a Cauchy surface since they do cross the event shell coming in. They would  be  asymptotic to the inner horizon on the other side except that  there is no matter inside, Kerr is singular in the middle and these rays do not connect.

There are many  reasons why I never believed that  Penrose  proved that black holes must be singular, \eg
\begin{enumerate}
    \item There are no trapped surfaces inside the inner horizon. One can always move outwards from any point inside this,  \eg $\bf k_+$. Even  time-like geodesics can do so. If a ray travels down the rotation axis from outside it can end up asymptotic to the inner horizon on the opposite side.  This is how I  proved  in 1963 that Kerr has two horizons. A graduate student, Alex Goodenbour,  recalculated this in 2021 with the same results.
    \item  It is not good enough to show that there is a FALL that is normal to the original trapped surface. It could finish tangential to the inner horizon rather than the interior of the central body. This has to be shown to be impossible, not just assumed or ignored.
    \item It is very much a "do it yourself" paper where the reader is supposed to prove the more difficult parts. Some of these may not even be true.
\end{enumerate}

Suppose we have an real  star that is spinning fast but is also shrinking and  is on the verge of forming a black hole. There will be a shell surrounding it that is difficult, but not impossible,  to escape from. As the star contracts further this will become harder and harder until an event shell  forms. Radial light rays that come down the axis through the North pole will pass through the star and die asymptotically on the inner horizon on the opposite side.

I cannot prove this because there is no agreed solution for the metric inside the star. What has been shown by example  is that FALL's can exist asymptotic to event horizons. All would be black hole singularity theorems must prove that this is impossible. This has not been done and so all the proofs of the various singularity theorems are incomplete. they always were since nobody could prove that FALL's imply singularities.

In desperation, many will say that  Kerr is just a special case. This is specious. Specific counterexamples are the standard way to disprove general claims in both Mathematics and Physics. The author does not have to prove that   any "apparent" inner horizon must have an asymptotic FALL but this is almost certainly true and is probably what the Penrose paper is trying to prove. 

Finally, what do I believe happens to a real collapsing neutron star? Suppose its mass  $m$  is $5M_\odot$,  larger than the Chandrasekhar limit for nonrotating ones, and that its radius is close to the Schwarzschild limit, $2m$.   It could be rotating so fast that the a-parameter is greater than m and no event horizon can form, so we will  assume that $a<m$ but close to it. Inside the star there will  be an equilibrium between the spin, pressure, and Newtonian type forces\footnote{Astrophysicists may say, "What is the equation of state?". This  is still a work in progress  for a neutron star.}.
As one moves away from the star, the spin forces  will drop faster than the  "Newtonian" ones. Very high initial velocities will therefore be needed to escape from S, even though the spin helps greatly near its surface.

If extra matter  joins the star, \eg  from the initial supernova, then  further collapse will probably take place. Unless the angular momentum to mass ratio rises very rapidly  the  outer shell will   become impenetrable and a black hole will form. 
This is the moment when the singularity is believed to appear, according to all true believers. {\it "No singularity before event shell forms, inevitable singularity afterwards!"}
What the author has tried to show is that there is no reason why this should happen.  Centrifugal forces will always dominate in the end as the radius of the body decreases. That is just Physics. The Kerr metric shows that there will be a region between the event shell and the central body where an eagle can fly if it flaps its wings hard enough. It will, of course, notice the outside universe spinning very quickly around it. It  may also have a problem with the radiation building up between the star and inner horizon.

This all ignores the "maximal extensions" of the exact solutions.  They are oddities with no physical significance and would require generating masses inside each inner section. They prove nothing more than  the original empty Kerr or Schwarzschild . They cannot be generated in the infinite past or future.  Furthermore, they do not counter the examples I have given here. Remember, one counterexample kills a universal claim.

In conclusion, I have tried to show that whatever the Penrose and Hawking theorems prove  has nothing to do with  Physics breaking down and singularities appearing. Of course, it is impossible to prove that these cannot exist, but it is extremely unlikely and goes against known physics.

\section{Appendix}
The  purpose of this section is to calculate the contravariant PNV's in Boyer-Lindquist and then Kerr-Schild coordinates.
The cyclic coordinates, $(t.\phi)$ are the only ones that change in any application of Papapetrou's theorem, neither $r$ nor $\theta$.
The suffix $s$ is used to show that the coordinates and the metric, $K_S$, correspond to those used for Schwarzschild. 
The transformation from Kerr to Papapetrou, \ie Boyer-Lindquist, coordinates is
\begin{subequations}
\beq       dt_s = dt - \frac{2mr}{\Delta} dr ,\qquad
        d\phi_s = d\phi + \frac{a}{\Delta} dr,\qquad r_s = r,\qquad \theta_s = \theta.
\eeq
\beq   
         \Delta = r^2 - 2mr + a^2,  \quad \Sigma = r^2 + a^2 \cos^2\theta. \
\eeq
\end{subequations}  
The only partial derivative operator that changes is $\partial_r$,
\begin{equation}  \label{E:Pdo}  (\partial_{r_s},\;\partial_{\theta_s},\;\partial_{\phi_s},\; \partial_{t_s}) = (\partial_{r} + \frac{2mr}{\Delta}\partial_{t} - \frac{a}{\Delta}\partial_{\phi} ,\:\partial_{\theta},\:\partial_{\phi}, \:\partial_{t})
\end{equation}
The covariant form of the metric, $ds^2$, is  a sum of squares of a suitably weighted orthonormal basis,
\begin{multline}\label{E:Blc}   
       \qquad ds^2 = \frac{\Sigma} {\Delta}dr^2   -\frac{\Delta}{\Sigma}\left(dt_s + a \sin^2\theta \,d\phi_s \right)^2,\qquad\\+ \Sigma d\theta^2  +  \frac{sin^2\theta}{\Sigma}\bigl( (r^2+a^2)d\phi_s + adt_s \bigr)^2
\end{multline}
The contravariant metric is a similar  sum of squares of the orthogonal tetrad,
\beq\label{E:Blcon}
\begin{split}
    \qquad     g^{\mu\nu}\partial_\mu\partial_\nu =  \frac {\Delta}{\Sigma}   {\partial_{r_s}}^2      &-\frac{1}{\Delta\Sigma}\left((r^2+a^2 )\partial_{t_s} - a\partial_{\phi_s}  \right)^2 \\
    &+ \frac {1}{\Sigma} \partial_{\theta_s^2 }
     +  \frac{1}{\Sigma\sin^2\theta } 
    \bigl(\partial_{\phi_s}  - a\sin^2\theta\partial_{t_s} \bigr)^2 ] \qquad
\end{split}
\eeq
Light rays are only defined up to a multiplicative constant which can depend on the ray, and so we will remove any overall factors when suitable. 

From \eqref{E:ks1}   the  PNV $k_-$ (the original $k$ in Kerr) is 
\beq\label{E:BLk}
        k_- =  (dt_s + a sin^2\theta d\phi_s) + (\Sigma\Delta^{-1})dr
\eeq
which  a combination of the first two terms in \ref{E:Blc}. 
Because the Boyer-Lindquist metric,  $K_S$, is invariant under the inversion ($t_s\rightarrow -t_s, \, \phi_s\rightarrow -\phi_s$),  the second PNV is  the other root of the first two terms in \eqref{E:Blc} and \eqref{E:Blcon}. In the original Kerr-Schild coordinates,
\begin{align}
         k_\pm &= \mp\Sigma dr + [\Delta(dt + a sin^2\theta d\phi)  +( -2mr + a^2 \sin^2 \theta)dr]\\
        {\bf k_\pm} &=  \mp(\Delta\partial_{r} + 2mr\partial_{t} - a\partial_{\phi}) + ((r^2+a^2 )\partial_{t} -a\partial_{\phi}   )   
\end{align}\label{}
 The contravariant version of the original PNV, $\bf k = k_-$,  is simpler than in its covariant form,
\begin{equation}
    {\bf k} = \partial_t -  \partial_r \quad\Longrightarrow \quad\frac{d r}{dt} = -1.
\end{equation}
 The  other  two  variables, $\phi$ and $\theta$, are constant along this PNV. The second PNV, $\bf k^*$, is more complicated,
Using $t$ as the best physical parameter along these rays
\beq
  \frac{dr}{dt} =   \frac{r^2 - 2mr + a^2}{r^2 + 2mr + a^2},\qquad \frac{d\phi}{dt} =  - \frac{2a}{r^2 + 2mr + a^2}.
\eeq
which shows that  $\bf{k}_+$ points inwards between the horizons but outwards elsewhere. lies on each event horizon when $\Delta(r) = 0$ and is  also parallel to  its  normal there.
 From  section~\ref{S:Affine},   whenever ${\bf \partial}_t$ is a Killing vector on a light ray then any affine parameter on the ray is an exponential function of $t$. For Kerr, it approaches a constant as $t\longrightarrow +\infty$ on the inner horizon  or as $t\longrightarrow -\infty$ on the outer horizon. These affine parameters can be chosen to be constant on each horizon so that $a$ is a smooth function throughout.

\section*{Acknowledgements}
I would like to thank Dr Chris Stevens and Alex Goudenbour of Canterbury University,  for many many useful discussions during this research.

\end{document}